\documentclass[aps,pra,preprint,superscriptaddress,showpacs]{revtex4-1}

\usepackage{graphicx}
\usepackage[english]{babel}
\addto\captionsenglish{}
\addto\captionsenglish{}
\usepackage{amsmath,amssymb,amsfonts,latexsym,amsbsy}
\usepackage{gensymb}
\usepackage[colorlinks=true,linkcolor=blue,citecolor=blue,urlcolor=blue]{hyperref}
\usepackage{xcolor}
\usepackage{lineno}

\begin{document}


\title{Two-level systems and growth-induced metastability in hydrogenated amorphous silicon}


\author{M. Molina-Ruiz}
\email[Corresponding author: ]{manelmolinaruiz@gmail.com}
\affiliation{Department of Physics, University of California Berkeley, Berkeley, CA 94720, USA}

\author{H. C. Jacks}
\altaffiliation{Present address: California Polytechnic University, Physics Department, San Luis Obispo, CA, 93407}
\affiliation{Department of Physics, University of California Berkeley, Berkeley, CA 94720, USA}

\author{D. R. Queen}
\altaffiliation{Present address: Northrop Grumman Corp., Linthicum, MD}
\affiliation{Department of Physics, University of California Berkeley, Berkeley, CA 94720, USA}

\author{Q. Wang}
\affiliation{National Renewable Energy Laboratory, Golden, Colorado 80401, USA}

\author{R. S. Crandall}
\affiliation{National Renewable Energy Laboratory, Golden, Colorado 80401, USA}

\author{F. Hellman}
\affiliation{Department of Physics, University of California Berkeley, Berkeley, CA 94720, USA}


\date{\today}

\begin{abstract}
Specific heat measurements from 2 to 300 K of hydrogenated amorphous silicon prepared by hot-wire chemical vapor deposition show a large excess specific heat at low temperature, significantly larger than the Debye specific heat calculated from the sound velocity. The as-prepared films have a Schottky anomaly that is associated with metastable hydrogen in the amorphous network, as well as large linear and excess cubic term commonly associated with tunneling two-level systems in amorphous solids. Annealing at 200 \degree C, a temperature that enables hydrogen mobility but not evaporation, irreversibly reduces the heat capacity, eliminating the Schottky anomaly and leaving a reduced linear heat capacity. A non-monotonic dependence on growth temperature and H content is observed in all effects, except for sound velocity, which suggests that the tunneling two-level systems and the Schottky anomaly are associated with atomic hydrogen and require low density regions to form, while sound velocity is associated with the silicon network and improves with increasing growth temperature.
\end{abstract}


\maketitle
\section{Introduction}
The low temperature properties of amorphous insulating solids are dominated by low energy excitations not normally found in crystalline solids~\cite{Zeller1971}. The specific heat, $C_P(T)$, of these solids is dominated by vibrational modes that are well described by the Debye model, which yields a cubic term, $c_DT^3$. These low energy excitations lead to an excess low T specific heat typically characterized by a linear term, $c_1T$~\cite{Pohl1981}. The linear part of the excess specific heat is described well by the standard tunneling model (STM) for two-level systems (TLS), which assumes that there is a finite tunneling probability between neighboring states in the disordered energy landscape~\cite{Anderson1972,Phillips1972,Phillips1987}. It is thought that these states correspond to groups of atoms that tunnel between energetically similar configurations.

The STM also explain other low temperature glassy properties, such as internal friction, $Q^{-1}(T)$, measurements in which the TLS lead to mechanical loss at low temperature.  Remarkably, the magnitude of $Q^{-1}(T)$ at low T (0.1 - 10 K) has a nearly universal value between 10$^{-3}$ and 10$^{-4}$, a result which has provoked much discussion over the decades without resolution, although a recent publication has suggested an explanation~\cite{Carruzzo2020WhyTemperatures}. Several decades ago, Phillips predicted that amorphous solids must have an open structure with low coordination for TLS to occur, and that it might be possible to depress TLS in tetrahedrally bonded systems such as amorphous silicon ($a$-Si) where the tetrahedral bonding over-constrains the atoms~\cite{Phillips1973}, but this prediction had not been tested. Recent internal friction and specific heat experiments showed that $a$-Si, without hydrogen, has a TLS density which ranges from vanishingly small to relatively large, determined by growth conditions, with a dependence on density that suggests that TLS originate in low density regions of the material~\cite{Zink2006,Queen2013,Liu2014,Queen2015JNCS}. Additionally, earlier work on hydrogenated amorphous silicon ($a$-Si:H) prepared by hot-wire chemical vapor deposition (HWCVD) was the first material found to have orders of magnitude fewer TLS than any other amorphous solids~\cite{Liu1997}. This work was followed by measurements of variably low TLS in tetrahedrally-bonded $a$-Si, $a$-Ge, $ta$-C, and their hydrogenated counterparts with and without secondary dopants, where it was suggested that TLS are sensitive to the existence of local underconstrained regions associated with structural inhomogeneities~\cite{Liu1998,Liu2002}. In particular, in these internal friction measurements on $a$-Si:H, it was thought that hydrogen was critical to removing TLS, but hydrogen proved to be an unreliable growth parameter because of the difficulty in controlling the $a$-Si structure and the hydrogen distribution, which depend on growth temperature~\cite{Yang2010}, and the subsequent work on $a$-Si, without H, showed that H was in fact not necessary to eliminate TLS~\cite{Queen2013,Liu2014,Queen2015JNCS}.

In addition to the linear heat capacity term, amorphous insulating solids also generally contain a higher-order excess specific heat term which theories try to explain, such as  the theory of random fluctuating transverse elastic constants~\cite{Schirmacher2006ThermalPeak}, the phonon-saddle transition approach~\cite{Grigera2003PhononLiquids}, and the soft-potential model (SPM)~\cite{ilyin1987parameters,Parshin1993SoftGlasses}. The latter is one of the most often considered approaches and has achieved good agreement with some experimental results~\cite{Buchenau1991AnharmonicGlasses,Ramos1998BeyondModel,Ramos2004AreDifferent}. The SPM postulates the coexistence of acoustic phonons with quasilocalized soft low-frequency modes. These modes add a $c_5T^5$ term to the cubic $c_DT^3$ (Debye) and linear $c_1T$ (TLS) terms.

Despite the success of the SPM in explaining the excess specific heat for several glasses, it has been shown in our previous work to fail when applied to $a$-Si~\cite{Queen2013}, where $c_5 = 0$ for all samples. Good fits for $C_P(T)$ for $a$-Si were achieved using the form $C_P = c_1T + c_3T^3$, where $c_3 = c_D + c_{ex}$, and $c_{ex}$ is a cubic excess specific heat term that systematically exceeds $c_D$ derived from measured elastic coefficients, and has no derivation in the standard tunneling model. In $a$-Si prepared by vapor deposition processes, $c_1$ and $c_{ex}$ systematically change more than two orders of magnitude with growth conditions, while $c_D$ varies by less than 20\%; the excess cubic term, $c_{ex}$, is shown to correlate over two decades with the linear term, $c_1$, and to correlate with the magnitude of the low T internal friction, strongly suggesting a common origin in the TLS~\cite{Queen2015JNCS}.

Because these tetrahedrally-bonded amorphous materials, including $a$-Si:H, can only be made as thin films by vapor deposition, measurements of TLS require techniques sensitive to the low mass of a thin film, such as nanocalorimetry. In the present work, we investigate the specific heat, density, and sound velocity of HWCVD $a$-Si:H films prepared at different growth temperatures yielding different H concentrations, both in their as-deposited and annealed states, and focus on the role that hydrogen plays in the formation of TLS in $a$-Si:H and the deviations from the TLS model that we observe.

\section{Experimental}
Device quality (low dangling bond density) $a$-Si:H thin films were prepared by HWCVD using pure silane, at growth rates between 1 to 3 nm/s, and substrate temperatures, $\textrm{T}_\textrm{S}$, of 300, 370, 430, and 470 \degree C~\cite{Mahan1991}. Higher $\textrm{T}_\textrm{S}$ yields films with systematically lower hydrogen content (from 9 to 3 at.\% H). Films were grown on membrane-based nanocalorimeters~\cite{Queen2009} as well as other substrates for other characterizations. Film thicknesses were measured with a KLA-Tencor Alpha-Step IQ profilometer. The atomic density of the films, $n_{at}$, was determined by Rutherford backscattering spectrometry in combination with profilometry measurements. Oxygen resonant scattering was used to probe for oxygen in the films; the only oxygen detected was limited to a thin surface layer. Hydrogen forward scattering was used to measure H content, and in particular to show that annealing at 200 and 300 \degree C, which allows hydrogen to equilibrate and redistribute, did not remove hydrogen from the films. Longitudinal sound velocities were measured at room temperature by a picosecond ultrasonic pump-probe technique~\cite{Lee2005}, with annealed films showing no significant change from their as-prepared counterparts. Room temperature thermal conductivity was measured using time-domain thermoreflectance and found to be comparable to previous HWCVD $a$-Si:H films~\cite{Yang2010}, but lower than the single high conductivity film reported in Ref.~\citenum{Liu2009}.

Specific heat, $C_{P}(T)$, was measured from 2 to 300 K using a microfabricated thin-film nanocalorimeter~\cite{Queen2009}. Sample area was defined lithographically after deposition and hence has very small error bar. The sample volume uncertainty was therefore determined by thickness uncertainty. The uncertainty in $C_{P}(T)$ is a combination of 5\% uncertainty in the background addenda~\cite{Queen2009} and 2\% uncertainty in sample thickness. Samples grown at 300, 370, and 430 \degree C were measured in their as-prepared and annealed states. Anneals were done in-situ alternately to specific heat measurements, in high vacuum for 10$^4$ seconds using the nanocalorimeter sample heater. The sample grown at 470 \degree C was measured and kept in its as-prepared state to verify that samples were stable throughout the duration of the study. After the first $C_{P}(T)$ measurement, samples were annealed at $\textrm{T}_\textrm{A}$ = 200 \degree C, then fast-cooled (quenched) at $\sim 10^{4}$ K/s, measured, annealed again at the same $\textrm{T}_\textrm{A}$, then slow-cooled at $\sim 5$ K/min and re-measured. This series was repeated for $\textrm{T}_\textrm{A}$ = 300 \degree C (for the samples grown at 370 and 430 \degree C). The fast cooling rate was achieved by turning off the sample heater power while keeping the thermal bath at 4.2 K; the slow cooling required software control to ensure a constant rate. The quenching preserves the high T equilibrated state, including any electronic or bonding defects, while slow cooling allows for further mobility and quasi-equilibration of the system at lower temperatures~\cite{Street1991}.

\section{Results and discussion}
\begin{figure}
\centering
\includegraphics{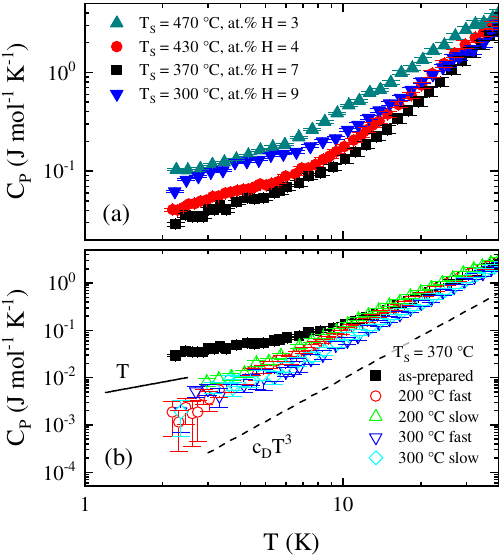}
\caption{\label{figure1}(a) Specific heat of four $a$-Si:H films grown at various temperatures in their as-prepared state. (b) Specific heat of $a$-Si:H film grown at $\textrm{T}_\textrm{S}$ = 370 \degree C in the as-prepared and annealed states, with $\textrm{T}_\textrm{A}$ = 200 or 300 \degree C. After annealing, the film was either fast- or slow-cooled as described in the text. The cubic specific heat term $c_DT^3$ (dashed line) is calculated from the measured sound velocity, which is weakly dependent on growth temperature and does not change upon annealing. The line labeled T is shown for comparison to a linear temperature dependence.}
\end{figure}

Figure~\hyperref[figure1]{\ref*{figure1}(a)} shows $C_{P}(T)$ of the four $a$-Si:H samples in their as-prepared state. Notably, $C_{P}(T)$ is not monotonic with growth temperature (nor with H content); instead the samples with the highest and lowest growth temperatures have specific heats significantly larger than those of the two intermediate growth temperatures. Figure~\hyperref[figure1]{\ref*{figure1}(b)} shows the effect of various anneals for the sample grown at 370 \degree C (7 at.\% H); similar results were found for the other samples upon annealing. The low T Debye specific heat, $c_{D}$, due to the phonon contribution (dashed line) was calculated from the measured longitudinal sound velocity and the  relationship $v_{t}=(0.56 \pm 0.05) v_{l}$ between the transverse and longitudinal sound velocities verified for many amorphous materials~\cite{Berret1988}, including our own $a$-Si grown at different temperatures~\cite{Queen2013}. We note that the longitudinal sound velocity, $v_l$, increases by $\sim15\%$ with $a$-Si:H growth temperature (see Table~\ref{table1}), as previously observed for $a$-Si~\cite{Queen2013}. The low temperature (below 10 K) $C_{P}(T)$ is significantly reduced after the first anneal, changes little with further annealing, and remains significantly larger than $c_D$, which shows no change on annealing. The large change in $C_{P}(T)$ on annealing, along with the lack of change in sound velocity, is a strong indication that the change in $C_{P}(T)$ is due to H rearrangement. We have observed no magnetic field dependence for $C_{P}(T)$ in these $a$-Si:H samples, indicating that electronic states are not responsible for the changes in $C_{P}(T)$~\cite{Queen2015EPL}. No cooling rate dependence was observed, indicating that metastable states, if present after annealing, equilibrate even during fast cooling. The differences in $C_{P}(T)$ between annealed states at different temperatures are negligible, so we focus our analysis on the differences between samples in as-prepared and 200 \degree C annealed states.

\begin{table*}
\caption{\label{table1}Summary of data: sample number (subscript `A' denotes the annealed states at 200 \degree C), growth temperature $\textrm{T}_\textrm{S}$, atomic hydrogen percentage at.\% H, thickness $t$, total atomic density $n_{at}$, longitudinal sound velocity $v_l$, excess specific heat $T^{3}$ term $c_{ex}$ calculated from $c_{ex} = c_3 - c_D$ [the cubic specific heat term, $c_3$, is obtained from low T fits using Eq.~\ref{equation4}, the Debye specific heat, $c_D$, is calculated from sound velocity, $v_{l}$, measurements, where $c_D = (12\pi^{4}/5) N_{A} k_{B} (T/T_{D})^{3}$, $T_D = (\hbar/k_{B}) (6 \pi^2 n_{at} {v_D}^3)^{1/3}$ and $v_D = (1/3{v_l}^{-3} + 2/3{v_t}^{-3})^{-1/3}$], TLS density $n_{0}$ calculated from Eq.~\ref{equation2}, number of systems $N$ (shown in mol$^{-1}$ of sample and per total number of atoms in \%), and energy splitting $\Delta_0$ for the Schottky anomaly calculated from Eq.~\ref{equation3}.}
\resizebox{\textwidth}{!}{
\begin{tabular}{c c c c c c c c c c c} \hline \hline
	No. & $\textrm{T}_\textrm{S}$ & at.\% H & $t$ & $n_{at}$ & $v_l$ & $c_{ex}$ & $n_0$ & $N$ & $N$ & $\Delta_0$ \\
 	 &  &  &  & x\,10$^{22}$ & x\,10$^{3}$ & x\,10$^{-6}$ & x\,10$^{47}$ & x\,10$^{21}$ &  &  \\
	 & (\degree C) & (\%) & (nm) & (atoms cm$^{-3}$) & (m s$^{-1}$) & (J mol$^{-1}$ K$^{-4}$) & (J$^{-1}$ m$^{-3}$) & (mol$^{-1}$) & (\%) & (K) \\ \hline
	1 & 300 & 9.0$\pm$0.5 & 79.2 & 4.8$\pm$0.2 & 7.1$\pm$0.3 & 62.1$\pm$18.0 & 33.5$\pm$4.9 & 9.6$\pm$1.4 & 1.6$\pm$0.2 & 9.7$\pm$0.3 \\
	1$_\textrm{A}$ & 300 & 9.0$\pm$0.5 & 79.2 & 4.8$\pm$0.2 & 7.1$\pm$0.3 & 128.9$\pm$7.8 & 5.6$\pm$0.7 & -- & -- & -- \\
	2 & 370 & 6.5$\pm$0.5 & 84.3 & 5.2$\pm$0.3 & 7.7$\pm$0.3 & 32.7$\pm$10.5 & 20.9$\pm$2.6 & 2.1$\pm$0.6 & 0.4$\pm$0.1 & 7.7$\pm$0.8 \\
	2$_\textrm{A}$ & 370 & 6.5$\pm$0.5 & 84.3 & 5.2$\pm$0.3 & 7.7$\pm$0.3 & 77.1$\pm$2.6 & 2.9$\pm$0.3 & -- & -- & -- \\
	3 & 430 & 4.0$\pm$0.5 & 88.6 & 5.0$\pm$0.1 & 7.9$\pm$0.3 & 62.9$\pm$5.6 & 23.8$\pm$1.5 & 3.7$\pm$0.3 & 0.6$\pm$0.1 & 7.7$\pm$0.2 \\
	3$_\textrm{A}$ & 430 & 4.0$\pm$0.5 & 88.6 & 5.0$\pm$0.1 & 7.9$\pm$0.3 & 120.3$\pm$4.6 & 3.4$\pm$0.2 & -- & -- & -- \\	
	4 & 470 & 3.0$\pm$0.5 & 78.7 & 4.4$\pm$0.1 & 8.1$\pm$0.3 & 136.4$\pm$13.4 & 57.3$\pm$2.8 & 7.1$\pm$0.8 & 1.2$\pm$0.1 & 4.4$\pm$0.8 \\ \hline \hline
\end{tabular}
}
\end{table*}

\begin{figure}
\centering
\includegraphics{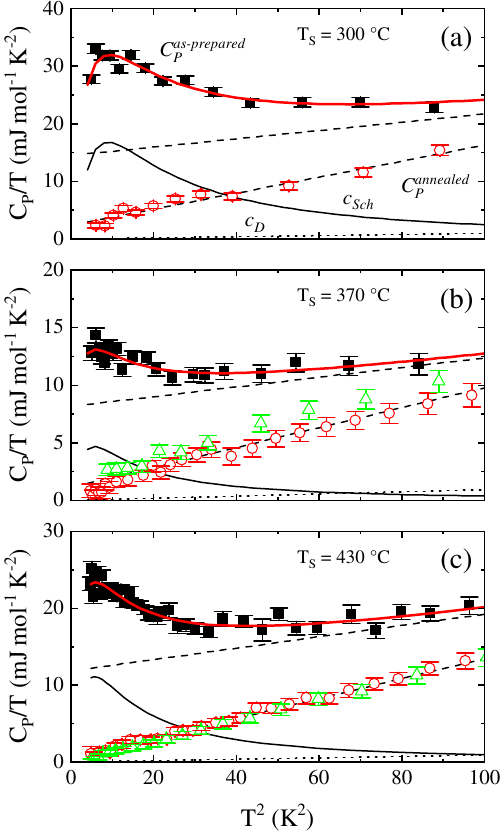}
\caption{\label{figure2}Specific heat of $a$-Si:H films grown at (a) 300 \degree C, (b) 370 \degree C, and (c) 430 \degree C plotted as $C_{P}/T$ versus $T^{2}$ in as-prepared (black squares), and 200 \degree C fast-cooled (red circles), and 200 \degree C slow-cooled (green triangles) annealed states. The red line is a fit to the as-prepared data $C_{P}(T)$ (Eq.~\ref{equation4}), the black line is the $c_{Sch}$ term (Eq.~\ref{equation3}), the dotted line is the $c_{D}$ term in both states, and the dashed lines correspond to the specific heat (Eq.~\ref{equation4} with $c_{Sch}=$ 0) in the as-prepared $C_{P}^{as-prepared}$, and the annealed $C_{P}^{annealed}$ states. For clarity, only the lines in plot (a) have been labeled.}
\end{figure}

Figure~\ref{figure2} shows $C_{P}(T)$ plotted as $C_{P}/T$ vs $T^{2}$ for the samples grown at 300, 370, and 430 \degree C. The specific heat signature for these samples is described by a large excess $C_{P}(T)$ in the as-prepared state, which includes a sub-linear temperature dependence (down to 2 K) that disappears upon annealing. The excess $C_{P}(T)$ in the as-prepared film must have a peak at or below 2 K, the lowest measured temperature, since $C_{P} \rightarrow 0$ as $T \rightarrow 0$ in order to have a finite entropy at $T = 0$ K. The simplest physical explanation for such a peak is a Schottky anomaly associated with two-level systems (which may or may not be due to tunneling) with a specific energy splitting $\Delta_0=E_0/k_B$.

The specific heat $C_{P}(T)$ of a distribution of two-level systems $n(\Delta)$ is given by
\begin{equation}
\text{\footnotesize$C_{P}(T)={k_{B}}^2\int_{0}^{\infty}n(\Delta)\bigg\{\left(\frac{\Delta}{T}\right)^2\frac{\textrm{exp}(-\Delta/T)}{[1+\textrm{exp}(-\Delta/T)]^2}\bigg\}d\Delta$}
\label{equation1}
\end{equation}
where $\Delta=E/k_B$ is the energy splitting of the two-level systems, $n(\Delta)$ the density of states for two-level systems with a particular $\Delta$, $k_{B}$ the Boltzmann constant, and $T$ the temperature. Note that Eq.~\ref{equation1} is valid both for the tunneling two-level systems (TLS) and for any other type of two-level systems. In the standard tunneling model, $n(\Delta)$ is taken as constant over a wide energy range due to the uniform distribution of tunnel barrier heights, so that $n(\Delta)\sim n_{0}$. Under these conditions, the solution for Eq.~\ref{equation1} yields the canonical linear specific heat term due to tunneling states~\cite{Anderson1972,Phillips1972}:
\begin{equation}
\text{\footnotesize$C_{P}=\frac{\pi^2}{6}{k_{B}}^2n_{0}\frac{N_{A}}{n_{at}}T=c_{1}T$}
\label{equation2}
\end{equation}
where $n_{0}$ is the density of TLS per unit energy and unit volume (J$^{-1}$m$^{-3}$), $n_{at}$ the total atomic density, and $N_{A}$ is Avogadro's number. If instead $n(\Delta)$ is considered to be a delta function, $N\delta(\Delta_0)$, to represent $N$ two-level systems per mol of sample with a specific energy splitting, $\Delta_0$, and assuming no degeneracy, from Eq.~\ref{equation1} we obtain the specific heat for a Schottky anomaly
\begin{equation}
\text{\footnotesize$c_{Sch}=Nk_{B}\left(\frac{\Delta_0}{T}\right)^2\frac{\textrm{exp}(-\Delta_0/T)}{[1+\textrm{exp}(-\Delta_0/T)]^2}$}
\label{equation3}
\end{equation}

The $a$-Si:H data presented in this work are well described by
\begin{equation}
\text{\footnotesize$C_{P}(T)=c_{1}T+c_{Sch}(T)+c_{3}T^3$}
\label{equation4}
\end{equation}
where the cubic term coefficient, $c_{3}$, is decomposed as $c_{3}=c_{D}+c_{ex}$, where $c_{D}$ is the Debye specific heat due to phonons, calculated from the sound velocity, and $c_{ex}$ is an excess specific heat $T^{3}$ term that is comparable in magnitude to $c_{D}$ but not explained by the standard tunneling model and likely associated with localized non-propagating modes~\cite{Stephens1973,Nakhmanson2000}. We have previously shown that in $a$-Si, $c_{ex}$ is proportional to $n_{0}$, which suggests a common structural origin~\cite{Queen2013}. The SPM, which adds an additional $c_5T^5$ term to account for the low temperature shoulder of the boson peak, does not seem to apply to $a$-Si:H; more specifically our attempts to fit our data to the low temperature tail of the boson peak using the SPM yield $c_5 \sim 0$, far too small in the SPM for the experimentally-determined $c_1$ linear term values.

Figure~\ref{figure2} shows fits to $C_{P}(T)$ below 10 K for the as-prepared and annealed states of different samples. Above $\sim 10$ K tunneling vanishes due to the appearance of thermally-activated processes, which are not described by the STM and, thus, Eq.~\ref{equation4} is no longer valid~\cite{Rau1995}. The red lines are fits to the as-prepared data using Eq.~\ref{equation4}; the dashed black lines show the canonical glassy fit ($C_{P}^{i}=c_{1}T+c_{3}T^{3}$) and the solid black lines show the Schottky term ($c_{Sch}$). Using multiple Schottky terms or degeneracy levels did not noticeably modify the fits, and in particular did not change the values of $n_{0}$ ($c_{1}$) or $c_{ex}$ ($c_{3}-c_{D}$). The $C_{P}(T)$ data after annealing are well-fit using the glassy expression Eq.~\ref{equation4} with $c_{Sch} = 0$. The sound velocity was found to be unchanged on annealing at 200 \degree C, so changes in $c_3$ are entirely due to changes in $c_{ex}$.

\begin{figure}
\centering
\includegraphics{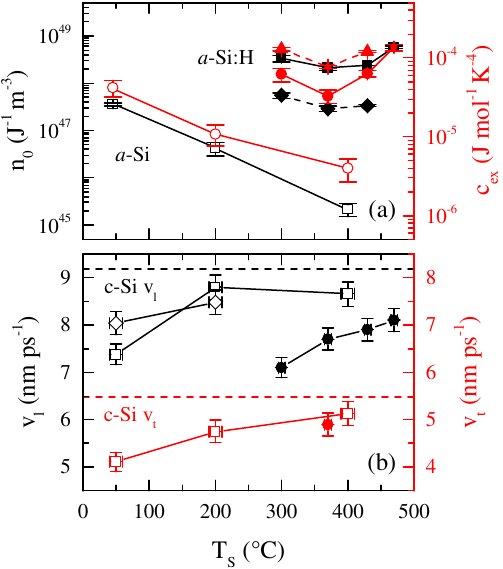}
\caption{\label{figure3}(a) Density of TLS $n_0$ (left black axis) and excess specific heat $T^3$ term $c_{ex}$ (right red axis) as a function of growth temperature, $\textrm{T}_\textrm{S}$, for both $a$-Si:H (closed symbols) and $a$-Si (open symbols from Ref.~\citenum{Queen2013}). The $n_0$ and $c_{ex}$ data for the as-prepared state is shown using solid lines, and black squares and red circles, respectively. The annealed state data for the $a$-Si:H samples is shown using dashed lines for $n_0$ (black rhombus) and $c_{ex}$ (red triangles). (b) Longitudinal (left black axis) and transverse (right red axis) sound velocity as a function of growth temperature for $a$-Si:H (closed symbols) and $a$-Si (open symbols from Ref.~\citenum{Queen2013}). Dashed lines indicate average longitudinal (black) and transverse (red) silicon (diamond structure) sound velocity.}
\end{figure}

The values of $n_{0}$ and $c_{ex}$ from the fits for as-prepared and annealed states are shown in Table~\ref{table1}, and plotted versus the growth temperature in Fig.~\ref{figure3} (closed symbols) along with the data from Ref.~\citenum{Queen2013} for $a$-Si (open symbols). Notably, annealing at 200 \degree C reduces $n_{0}$ by an order of magnitude in all the hydrogenated samples (black solid squares and diamonds), while $n_{0}$ of pure $a$-Si samples (black open squares) was not affected by thermal treatment (up to 200 \degree C), again an indication that $n_0$ of $a$-Si:H is significantly affected by H redistribution. As found in pure $a$-Si, $a$-Si:H films with larger $n_{0}$ also have larger $c_{ex}$ (red symbols), suggesting again that the structures responsible for TLS and non-propagating modes in $a$-Si:H are related to each other. The values of $n_{0}$ (and $c_{ex}$) are larger in $a$-Si:H than those for $a$-Si grown at similar temperatures. The dangling bond density, $n_{DB}$, in HWCVD $a$-Si:H samples is $\leq$ 10$^{16}$ cm$^{-3}$~\cite{Liu2002}, while for e-beam $a$-Si $n_{DB}\sim$ 10$^{19}$ cm$^{-3}$~\cite{Queen2013}. That the values of $n_0$ for $a$-Si:H are orders of magnitude higher than in $a$-Si samples grown at similar temperatures, shows there is no direct correlation between dangling bonds and TLS.  Additionally, the lack of proportionality between $n_0$ and H at.\% shows that H is not directly responsible for TLS, despite the fact that its presence enables TLS to occur in $a$-Si:H with far greater density than in $a$-Si; this point is central to this paper, and will be discussed further below.

In $a$-Si we observed that increasing the growth temperature reduces the density of TLS, and we suggested this is because higher growth temperature enables Si adatoms to find more stable sites during growth, leading to a denser and more energetically stable structure, which in turn leads to low TLS due to the absence of ``nearby'' similar energy wells that enable tunneling states to form~\cite{Queen2013,Queen2015JNCS}. This behavior, along with more recent as-yet unpublished data on the related effects of growth rate on atomic and TLS densities in $a$-Si, suggests that $a$-Si is being grown under conditions that cause it to be ultrastable (the most stable glasses are obtained when grown between 0.7 to 0.9 times the material glass transition~\cite{Ediger2017Perspective:Glasses}). In $a$-Si:H, the specific heat extracted values in both the as-prepared and annealed states (Fig.~\ref{figure3} and Table~\ref{table1}), show a non-monotonic behavior versus growth temperature, with a minimum for films grown at 370 \degree C. The enhanced stability is also related to increased atomic density, which in this case for $a$-Si:H is seen at 370 \degree C, matching with the lowest TLS density values.

The Schottky peak amplitude also shows the same minimum with growth T, as seen most clearly in Fig.~\ref{figure2} and in Table~\ref{table1}. When the separate fitting parameters $N$ and $\Delta_0$ are plotted, the minimum is clear in the number of Schottky systems $N$, but the energy gap $\Delta_0$ becomes monotonically smaller as $\textrm{T}_\textrm{S}$ increases. This effect may be a result of the limits of the fit, which is over a limited temperature range and assumes a single gap $\Delta_0$, or may be a real shift to lower $\Delta_0$ with higher $\textrm{T}_\textrm{S}$. We considered a wide range of fitting parameters, in which $N$ and $\Delta_0$ were either held constant or free; allowing them each to vary yields the best fitting results, without significantly modifying the other fit parameters ($c_1$ and $c_{ex}$).

We next consider the structural motifs that may give rise to the TLS in $a$-Si:H. Based on the nuclear magnetic resonance (NMR) work by Crandall~\cite{Wu1996}, we consider two different structural regions: (i) regions with high structural order regions, or network, containing very few H atoms, and (ii) lower structural order regions that give rise to defects and low density structures, including dangling bonds and nanovoids, where H atoms may cluster. We propose that TLS occur in the lower structural order regions, while sound waves are predominantly carried by the network. The sound velocity, $v_l$, linearly increases with growth temperature, showing a similar dependence as previously found for hydrogen free $a$-Si [see Fig.~\hyperref[figure3]{\ref*{figure3}(b)}], and suggesting that the network becomes more ordered (specifically, reduced bond angle disorder) with increasing growth temperature. By contrast, the atomic density, $n_{at}$, is non-monotonic in growth temperature, with a maximum at 370 \degree C and a strong reduction at both higher and lower $\textrm{T}_\textrm{S}$, indicating an increased amount of low density regions at both low and high $\textrm{T}_\textrm{S}$. That increased amount of low density regions in turn leads to a higher amount of TLS ($n_0$ and $c_{ex}$) as well as the traditional two-level systems ($c_{Sch}$, associated with a single $\Delta_0$). We suggest that lower density regions contain underconstrained Si atoms and H, which lead to TLS and to the Schottky anomaly. Most strikingly, as seen in Fig.~\ref{figure3}, the presence of H causes high $n_0$ and $c_{ex}$, relative to $a$-Si, but these excess heat capacities are not proportional to the at.\% H (Table~\ref{table1}), and instead depend on the presence of both H and low density regions. Annealing redistributes H which leaves the sound velocity and atomic density unchanged but eliminates $c_{Sch}$, reduces $n_0$ and increases $c_{ex}$, although still leaving these larger than in $a$-Si. 

The local hydrogen environment in similar HWCVD $a$-Si:H samples shows that hydrogen is primarily found as bonded Si--H, and secondarily as molecular hydrogen (H$_2$), in both cases either clustered or isolated~\cite{Wu1996,Su2000,Reimer1981}. The NMR spectra change irreversibly upon annealing at $\textrm{T}_\textrm{A}$ comparable to that used here, which is notably below $\textrm{T}_\textrm{S}$ and show that the as-prepared state is metastable~\cite{Baugh1999}. At temperatures above the H mobility threshold ($\sim 100$ \degree C)~\cite{Street1991}, hydrogen diffuses away from clustered Si--H regions and increases the fraction of isolated Si--H bonds, where it is trapped via a mechanism of reconstructing Si--H bonds~\cite{Beyer1983}. The trapped states that occur when H diffuses to a locally deep potential well originate either from a weak Si--Si bond that forms a dangling bond plus a Si--H bond~\cite{Branz1998}, or a dangling bond that then forms a Si--H bond~\cite{Fujita1994KineticsSilicon}. Hydrogen atoms may also combine with other H to produce H$_2$~\cite{Baugh1999}.

Hydrogen molecules have been shown to lead to a Schottky anomaly below 5 K, attributed to the freezing of the rotational motion of ortho-H$_2$ trapped in the film~\cite{Graebner1985}. Annealing might increase the molecular H$_2$ concentration or might leave it unchanged but, in the present study, we find that annealing \textit{removes} the Schottky anomaly, showing that the Schottky anomaly is not associated with molecular H$_2$. Furthermore, we see no sign of the H$_2$ melting point in $C_P(T)$, unlike that observed in Ref.~\citenum{Graebner1985}, implying that clusters of H$_2$ are not large enough to be detected by these means. These observations show that any molecular hydrogen is present as isolated H$_2$ and is not the source of the Schottky anomaly. We instead suggest that the Schottky anomaly is associated with distinct H sites with slightly different energies near metastable weak Si--Si bonds, such as would be found in low density, disordered regions of the film. In support of this idea, we note that the percentage of the number of systems, $N$, with energy splitting $\Delta_0$ is comparable to but smaller than the at.\% H (Table~\ref{table1}).

To understand the dependence of $n_{0}$ and $c_{ex}$ on growth temperature, we first consider the films’ atomic density, $n_{at}$, shown in Table~\ref{table1}, since in our previous work on $a$-Si we showed that $n_{0}$ and $c_{ex}$ have a strong correlation with $n_{at}$~\cite{Queen2013}. Here in $a$-Si:H, $n_{at}$ shows the same dependence on growth temperature as $n_0$, $c_{ex}$ and $c_{Sch}$, suggesting a similar dependence of TLS density on atomic density, although for any given $n_{at}$ the TLS density, $n_0$, is at least 2 orders of magnitude larger in $a$-Si:H than in $a$-Si, an effect we attribute to the presence of H and the inherent differences between the growth techniques used. This optimum growth temperature behavior, resulting in the lowest $n_{0}$, $c_{ex}$, $c_{Sch}$ and highest $n_{at}$ values, is notably replicated by Urbach edge and $n_{DB}$ measurements on device quality HWCVD $a$-Si:H~\cite{Mahan1991}, techniques traditionally used to establish the film quality. It is similarly notable that the reduction in $n_{DB}$ in device quality $a$-Si:H requires H, but is not proportional to its content. The significant variation in atomic density with growth temperature cannot be explained solely in terms of network distortion, nor dangling bonds, but instead requires the presence of nanovoids or loosely bonded regions. As discussed by Street~\cite{Street1991a}, both high and low $\textrm{T}_\textrm{S}$ result in a more disordered structure; here we show that this disorder correlates with lower atomic and higher TLS densities. We note the strong and monotonic increase of sound velocity with $\textrm{T}_\textrm{S}$ showing that the network becomes more ordered, and it is only the low density regions where disorder increases. Notably, there is no change in $n_{at}$ or H content upon annealing, despite a reduction in $n_0$ and a vanishing $c_{Sch}$, showing that these excess specific heat terms are related to H in the low density regions, not to the presence of low density regions alone. At these relatively low annealing temperatures, annealing is known to result in H mobility, relocation and rebonding~\cite{Reimer1981,Beyer1983,Baugh1999,Branz1998}, which affect $n_{0}$ and $c_{Sch}$, as well as $c_{ex}$, which will be discussed below.

It is known that hydrogen lowers the energy barrier for breaking weak Si--Si bonds~\cite{Biswas1999}. Upon annealing, H diffuses from clustered Si--H to locally deep potential wells increasing the isolated Si--H bonds fraction, and perhaps the H$_2$ concentration. These results suggest that both the Schottky anomaly and the large $n_0$ in the as-prepared state are associated with clustered Si--H. The lack of correlation of at.\% H with TLS density, despite the necessity of H to produce the large TLS, is similar to the effects seen with elimination of dangling bonds in $a$-Si:H, where H is essential to create device quality electronic materials, but the necessary at.\% ranges from 1 to 20\%, depending on preparation details, and is uncorrelated with the reduction in $n_{DB}$.

After annealing, the $a$-Si:H samples presented in this work show a decoupling between $n_0$ and $c_{ex}$ not observed in $a$-Si, despite the strong correlation in the as-prepared state. The TLS density, $n_0$, decreases (solid to dashed black lines in Fig.~\ref{figure3}), while the excess specific heat $T^3$ term, $c_{ex}$, increases (solid to dashed red lines in Fig.~\ref{figure3}), despite both showing the same minimum in the highest atomic density films grown at 370 \degree C. This brings into question the simple explanation of $c_{ex}$ as being directly due to the structures that give rise to TLS. Instead, we propose that $c_{ex}$ is caused by low density regions, which give rise to the vibrational states detailed by Nakhmanson and Drabold~\cite{Nakhmanson2000,Nakhmanson2000a}. In these models, low energy vibrational states are found to occur in lower structural order regions that also accommodate TLS. Annealing-induced H redistribution decreases the TLS, by eliminating/reducing weak Si--Si bonds, but does not eliminate the underlying structural defects that caused the TLS (nanovoids or concomitant strained regions). The increase in $c_{ex}$ upon annealing is likely also associated with the same H diffusion process, causing a softening of the local modes localized in lower structural order regions.

\section{Conclusions}
A large excess specific heat is seen at low temperatures in as-prepared films of $a$-Si:H that show a Schottky anomaly in addition to the linear and cubic terms that are commonly seen in amorphous materials and attributed to tunneling two-level systems. The Schottky anomaly as well as the large value of $n_0$ decrease upon annealing at modest temperatures (200 \degree C), well below the growth temperature, with the former vanishing entirely. These annealing conditions are known to cause hydrogen to diffuse from a clustered Si--H to a more uniform distribution of isolated Si--H bonds.

The specific heat observations are suggested to be due to clustered atomic H in low density regions, which leads to both the two-level systems that produce the Schottky anomaly (H near weak Si--Si bonds) and, via depressed energy barriers, a greater probability of atomic rearrangements of Si structures that produce TLS. A strong correlation of $n_0$, $c_{ex}$, and $c_{Sch}$ with atomic density, which is not monotonic in growth temperature, and the lack of correlation with at.\% H, despite the critical role  that H plays in creating these specific heat effects, demonstrates that low density regions are essential to producing these effects, as was seen in pure $a$-Si. Atomic H thus greatly enhances the TLS compared to that seen in pure $a$-Si by lowering energy barriers, but the effects are not proportional to at.\% H, similar to the lack of proportionality between dangling bond density and at.\% H in device quality $a$-Si:H. The monotonic increase of the sound velocity with growth temperature also suggests that TLS originate in low density regions, and not in the network.

The nature of the cubic $c_{ex}$ term is not yet understood, but the correlation in the as-prepared state with $n_0$ (as was found in pure $a$-Si) implies that the structures responsible for its existence are related to those that produce the TLS density, $n_0$. The different effect of annealing on $n_0$ and $c_{ex}$ of $a$-Si:H points to an intriguing interpretation; the same structures lead to both, but atomic H is essential for $n_0$ and not for $c_{ex}$.
%
\begin{acknowledgments}
We thank E. Iwaniczko for preparation of the $a$-Si:H films; G. Hohensee and D. G. Cahill for sound velocity and thermal conductivity measurements; J. A. Reimer for helpful discussions. We thank the NSF DMR-0907724, 1508828, and 1809498 for support of this project.
\end{acknowledgments}
%
\bibliography{references}

\begin{thebibliography}{44}%
\makeatletter
\providecommand \@ifxundefined [1]{%
 \@ifx{#1\undefined}
}%
\providecommand \@ifnum [1]{%
 \ifnum #1\expandafter \@firstoftwo
 \else \expandafter \@secondoftwo
 \fi
}%
\providecommand \@ifx [1]{%
 \ifx #1\expandafter \@firstoftwo
 \else \expandafter \@secondoftwo
 \fi
}%
\providecommand \natexlab [1]{#1}%
\providecommand \enquote  [1]{``#1''}%
\providecommand \bibnamefont  [1]{#1}%
\providecommand \bibfnamefont [1]{#1}%
\providecommand \citenamefont [1]{#1}%
\providecommand \href@noop [0]{\@secondoftwo}%
\providecommand \href [0]{\begingroup \@sanitize@url \@href}%
\providecommand \@href[1]{\@@startlink{#1}\@@href}%
\providecommand \@@href[1]{\endgroup#1\@@endlink}%
\providecommand \@sanitize@url [0]{\catcode `\\12\catcode `\$12\catcode
  `\&12\catcode `\#12\catcode `\^12\catcode `\_12\catcode `\%12\relax}%
\providecommand \@@startlink[1]{}%
\providecommand \@@endlink[0]{}%
\providecommand \url  [0]{\begingroup\@sanitize@url \@url }%
\providecommand \@url [1]{\endgroup\@href {#1}{\urlprefix }}%
\providecommand \urlprefix  [0]{URL }%
\providecommand \Eprint [0]{\href }%
\providecommand \doibase [0]{http://dx.doi.org/}%
\providecommand \selectlanguage [0]{\@gobble}%
\providecommand \bibinfo  [0]{\@secondoftwo}%
\providecommand \bibfield  [0]{\@secondoftwo}%
\providecommand \translation [1]{[#1]}%
\providecommand \BibitemOpen [0]{}%
\providecommand \bibitemStop [0]{}%
\providecommand \bibitemNoStop [0]{.\EOS\space}%
\providecommand \EOS [0]{\spacefactor3000\relax}%
\providecommand \BibitemShut  [1]{\csname bibitem#1\endcsname}%
\let\auto@bib@innerbib\@empty
\bibitem [{\citenamefont {Zeller}\ and\ \citenamefont
  {Pohl}(1971)}]{Zeller1971}%
  \BibitemOpen
  \bibfield  {author} {\bibinfo {author} {\bibfnamefont {R.~C.}\ \bibnamefont
  {Zeller}}\ and\ \bibinfo {author} {\bibfnamefont {R.~O.}\ \bibnamefont
  {Pohl}},\ }\href {https://doi.org/10.1103/PhysRevB.4.2029} {\bibfield
  {journal} {\bibinfo  {journal} {Phys. Rev. B}\ }\textbf {\bibinfo {volume}
  {4}},\ \bibinfo {pages} {2029} (\bibinfo {year} {1971})}\BibitemShut
  {NoStop}%
\bibitem [{\citenamefont {Pohl}(1981)}]{Pohl1981}%
  \BibitemOpen
  \bibfield  {author} {\bibinfo {author} {\bibfnamefont {R.~O.}\ \bibnamefont
  {Pohl}},\ }\href {\doibase 10.1007/978-3-642-81534-8} {\emph {\bibinfo
  {title} {{Amorphous Solids Low Temperature Properties}}}},\ edited by\
  \bibinfo {editor} {\bibfnamefont {W.~A.}\ \bibnamefont {Phillips}},\ \bibinfo
  {series} {Topics in Current Physics}, Vol.~\bibinfo {volume} {24}\ (\bibinfo
  {publisher} {Springer-Verlag},\ \bibinfo {year} {1981})\ pp.\ \bibinfo
  {pages} {27--51}\BibitemShut {NoStop}%
\bibitem [{\citenamefont {Anderson}\ \emph {et~al.}(1972)\citenamefont
  {Anderson}, \citenamefont {Halperin},\ and\ \citenamefont
  {Varma}}]{Anderson1972}%
  \BibitemOpen
  \bibfield  {author} {\bibinfo {author} {\bibfnamefont {P.~W.}\ \bibnamefont
  {Anderson}}, \bibinfo {author} {\bibfnamefont {B.~I.}\ \bibnamefont
  {Halperin}}, \ and\ \bibinfo {author} {\bibfnamefont {C.~M.}\ \bibnamefont
  {Varma}},\ }\href {\doibase 10.1080/14786437208229210} {\bibfield  {journal}
  {\bibinfo  {journal} {Philos. Mag.}\ }\textbf {\bibinfo {volume} {25}},\
  \bibinfo {pages} {1} (\bibinfo {year} {1972})}\BibitemShut {NoStop}%
\bibitem [{\citenamefont {Phillips}(1972)}]{Phillips1972}%
  \BibitemOpen
  \bibfield  {author} {\bibinfo {author} {\bibfnamefont {W.~A.}\ \bibnamefont
  {Phillips}},\ }\href {\doibase 10.1007/BF00660072} {\bibfield  {journal}
  {\bibinfo  {journal} {J. Low Temp. Phys.}\ }\textbf {\bibinfo {volume} {7}},\
  \bibinfo {pages} {351} (\bibinfo {year} {1972})}\BibitemShut {NoStop}%
\bibitem [{\citenamefont {Phillips}(1987)}]{Phillips1987}%
  \BibitemOpen
  \bibfield  {author} {\bibinfo {author} {\bibfnamefont {W.~A.}\ \bibnamefont
  {Phillips}},\ }\href {http://dx.doi.org/10.1088/0034-4885/50/12/003}
  {\bibfield  {journal} {\bibinfo  {journal} {Rep. Prog. Phys.}\ }\textbf
  {\bibinfo {volume} {50}},\ \bibinfo {pages} {1657} (\bibinfo {year}
  {1987})}\BibitemShut {NoStop}%
\bibitem [{\citenamefont {Carruzzo}\ and\ \citenamefont
  {Yu}(2020)}]{Carruzzo2020WhyTemperatures}%
  \BibitemOpen
  \bibfield  {author} {\bibinfo {author} {\bibfnamefont {H.~M.}\ \bibnamefont
  {Carruzzo}}\ and\ \bibinfo {author} {\bibfnamefont {C.~C.}\ \bibnamefont
  {Yu}},\ }\href {\doibase 10.1103/PhysRevLett.124.075902} {\bibfield
  {journal} {\bibinfo  {journal} {Phys. Rev. Lett.}\ }\textbf {\bibinfo
  {volume} {124}},\ \bibinfo {pages} {075902} (\bibinfo {year}
  {2020})}\BibitemShut {NoStop}%
\bibitem [{\citenamefont {Phillips}(1973)}]{Phillips1973}%
  \BibitemOpen
  \bibfield  {author} {\bibinfo {author} {\bibfnamefont {W.~A.}\ \bibnamefont
  {Phillips}},\ }\href {https://doi.org/10.1007/BF00654457} {\bibfield
  {journal} {\bibinfo  {journal} {J. Low Temp. Phys.}\ }\textbf {\bibinfo
  {volume} {11}},\ \bibinfo {pages} {757} (\bibinfo {year} {1973})}\BibitemShut
  {NoStop}%
\bibitem [{\citenamefont {Zink}\ \emph {et~al.}(2006)\citenamefont {Zink},
  \citenamefont {Pietri},\ and\ \citenamefont {Hellman}}]{Zink2006}%
  \BibitemOpen
  \bibfield  {author} {\bibinfo {author} {\bibfnamefont {B.~L.}\ \bibnamefont
  {Zink}}, \bibinfo {author} {\bibfnamefont {R.}~\bibnamefont {Pietri}}, \ and\
  \bibinfo {author} {\bibfnamefont {F.}~\bibnamefont {Hellman}},\ }\href
  {\doibase 10.1103/PhysRevLett.96.055902} {\bibfield  {journal} {\bibinfo
  {journal} {Phys. Rev. Lett.}\ }\textbf {\bibinfo {volume} {96}},\ \bibinfo
  {pages} {055902} (\bibinfo {year} {2006})}\BibitemShut {NoStop}%
\bibitem [{\citenamefont {Queen}\ \emph {et~al.}(2013)\citenamefont {Queen},
  \citenamefont {Liu}, \citenamefont {Karel}, \citenamefont {Metcalf},\ and\
  \citenamefont {Hellman}}]{Queen2013}%
  \BibitemOpen
  \bibfield  {author} {\bibinfo {author} {\bibfnamefont {D.~R.}\ \bibnamefont
  {Queen}}, \bibinfo {author} {\bibfnamefont {X.}~\bibnamefont {Liu}}, \bibinfo
  {author} {\bibfnamefont {J.}~\bibnamefont {Karel}}, \bibinfo {author}
  {\bibfnamefont {T.~H.}\ \bibnamefont {Metcalf}}, \ and\ \bibinfo {author}
  {\bibfnamefont {F.}~\bibnamefont {Hellman}},\ }\href {\doibase
  10.1103/PhysRevLett.110.135901} {\bibfield  {journal} {\bibinfo  {journal}
  {Phys. Rev. Lett.}\ }\textbf {\bibinfo {volume} {110}},\ \bibinfo {pages}
  {135901} (\bibinfo {year} {2013})}\BibitemShut {NoStop}%
\bibitem [{\citenamefont {Liu}\ \emph {et~al.}(2014)\citenamefont {Liu},
  \citenamefont {Queen}, \citenamefont {Metcalf}, \citenamefont {Karel},\ and\
  \citenamefont {Hellman}}]{Liu2014}%
  \BibitemOpen
  \bibfield  {author} {\bibinfo {author} {\bibfnamefont {X.}~\bibnamefont
  {Liu}}, \bibinfo {author} {\bibfnamefont {D.~R.}\ \bibnamefont {Queen}},
  \bibinfo {author} {\bibfnamefont {T.~H.}\ \bibnamefont {Metcalf}}, \bibinfo
  {author} {\bibfnamefont {J.~E.}\ \bibnamefont {Karel}}, \ and\ \bibinfo
  {author} {\bibfnamefont {F.}~\bibnamefont {Hellman}},\ }\href {\doibase
  10.1103/PhysRevLett.113.025503} {\bibfield  {journal} {\bibinfo  {journal}
  {Phys. Rev. Lett.}\ }\textbf {\bibinfo {volume} {113}},\ \bibinfo {pages}
  {025503} (\bibinfo {year} {2014})}\BibitemShut {NoStop}%
\bibitem [{\citenamefont {Queen}\ \emph
  {et~al.}(2015{\natexlab{a}})\citenamefont {Queen}, \citenamefont {Liu},
  \citenamefont {Karel}, \citenamefont {Jacks}, \citenamefont {Metcalf},\ and\
  \citenamefont {Hellman}}]{Queen2015JNCS}%
  \BibitemOpen
  \bibfield  {author} {\bibinfo {author} {\bibfnamefont {D.~R.}\ \bibnamefont
  {Queen}}, \bibinfo {author} {\bibfnamefont {X.}~\bibnamefont {Liu}}, \bibinfo
  {author} {\bibfnamefont {J.}~\bibnamefont {Karel}}, \bibinfo {author}
  {\bibfnamefont {H.~C.}\ \bibnamefont {Jacks}}, \bibinfo {author}
  {\bibfnamefont {T.~H.}\ \bibnamefont {Metcalf}}, \ and\ \bibinfo {author}
  {\bibfnamefont {F.}~\bibnamefont {Hellman}},\ }\href {\doibase
  10.1016/j.jnoncrysol.2015.06.020} {\bibfield  {journal} {\bibinfo  {journal}
  {J. Non-Cryst. Solids}\ }\textbf {\bibinfo {volume} {426}},\ \bibinfo {pages}
  {19} (\bibinfo {year} {2015}{\natexlab{a}})}\BibitemShut {NoStop}%
\bibitem [{\citenamefont {Liu}\ \emph {et~al.}(1997)\citenamefont {Liu},
  \citenamefont {Pohl}, \citenamefont {Crandall},\ and\ \citenamefont
  {Jones}}]{Liu1997}%
  \BibitemOpen
  \bibfield  {author} {\bibinfo {author} {\bibfnamefont {X.}~\bibnamefont
  {Liu}}, \bibinfo {author} {\bibfnamefont {R.~O.}\ \bibnamefont {Pohl}},
  \bibinfo {author} {\bibfnamefont {R.~S.}\ \bibnamefont {Crandall}}, \ and\
  \bibinfo {author} {\bibfnamefont {K.~M.}\ \bibnamefont {Jones}},\ }\href
  {\doibase 10.1557/PROC-469-419} {\bibfield  {journal} {\bibinfo  {journal}
  {Mat. Res. Soc. Symp. Proc}\ }\textbf {\bibinfo {volume} {469}},\ \bibinfo
  {pages} {419} (\bibinfo {year} {1997})}\BibitemShut {NoStop}%
\bibitem [{\citenamefont {Liu}\ \emph {et~al.}(1998)\citenamefont {Liu},
  \citenamefont {Iwaniczko}, \citenamefont {Pohl},\ and\ \citenamefont
  {Crandall}}]{Liu1998}%
  \BibitemOpen
  \bibfield  {author} {\bibinfo {author} {\bibfnamefont {X.}~\bibnamefont
  {Liu}}, \bibinfo {author} {\bibfnamefont {E.}~\bibnamefont {Iwaniczko}},
  \bibinfo {author} {\bibfnamefont {R.~O.}\ \bibnamefont {Pohl}}, \ and\
  \bibinfo {author} {\bibfnamefont {R.~S.}\ \bibnamefont {Crandall}},\ }\href
  {https://doi.org/10.1557/PROC-507-595} {\bibfield  {journal} {\bibinfo
  {journal} {Mat. Res. Soc. Symp. Proc.}\ }\textbf {\bibinfo {volume} {507}},\
  \bibinfo {pages} {595} (\bibinfo {year} {1998})}\BibitemShut {NoStop}%
\bibitem [{\citenamefont {Liu}\ \emph {et~al.}(2002)\citenamefont {Liu},
  \citenamefont {Photiadis}, \citenamefont {Wu}, \citenamefont {Chrisey},
  \citenamefont {Pohl},\ and\ \citenamefont {Crandall}}]{Liu2002}%
  \BibitemOpen
  \bibfield  {author} {\bibinfo {author} {\bibfnamefont {X.}~\bibnamefont
  {Liu}}, \bibinfo {author} {\bibfnamefont {D.~M.}\ \bibnamefont {Photiadis}},
  \bibinfo {author} {\bibfnamefont {H.-D.}\ \bibnamefont {Wu}}, \bibinfo
  {author} {\bibfnamefont {D.~B.}\ \bibnamefont {Chrisey}}, \bibinfo {author}
  {\bibfnamefont {R.~O.}\ \bibnamefont {Pohl}}, \ and\ \bibinfo {author}
  {\bibfnamefont {R.~S.}\ \bibnamefont {Crandall}},\ }\href {\doibase
  10.1080/13642810208208541} {\bibfield  {journal} {\bibinfo  {journal}
  {Philos. Mag. B}\ }\textbf {\bibinfo {volume} {82}},\ \bibinfo {pages} {185}
  (\bibinfo {year} {2002})}\BibitemShut {NoStop}%
\bibitem [{\citenamefont {Yang}\ \emph {et~al.}(2010)\citenamefont {Yang},
  \citenamefont {Cahill}, \citenamefont {Liu}, \citenamefont {Feldman},
  \citenamefont {Crandall}, \citenamefont {Sperling},\ and\ \citenamefont
  {Abelson}}]{Yang2010}%
  \BibitemOpen
  \bibfield  {author} {\bibinfo {author} {\bibfnamefont {H.-S.}\ \bibnamefont
  {Yang}}, \bibinfo {author} {\bibfnamefont {D.~G.}\ \bibnamefont {Cahill}},
  \bibinfo {author} {\bibfnamefont {X.}~\bibnamefont {Liu}}, \bibinfo {author}
  {\bibfnamefont {J.~L.}\ \bibnamefont {Feldman}}, \bibinfo {author}
  {\bibfnamefont {R.~S.}\ \bibnamefont {Crandall}}, \bibinfo {author}
  {\bibfnamefont {B.~A.}\ \bibnamefont {Sperling}}, \ and\ \bibinfo {author}
  {\bibfnamefont {J.~R.}\ \bibnamefont {Abelson}},\ }\href {\doibase
  10.1103/PhysRevB.81.104203} {\bibfield  {journal} {\bibinfo  {journal} {Phys.
  Rev. B}\ }\textbf {\bibinfo {volume} {81}},\ \bibinfo {pages} {104203}
  (\bibinfo {year} {2010})}\BibitemShut {NoStop}%
\bibitem [{\citenamefont {Schirmacher}(2006)}]{Schirmacher2006ThermalPeak}%
  \BibitemOpen
  \bibfield  {author} {\bibinfo {author} {\bibfnamefont {W.}~\bibnamefont
  {Schirmacher}},\ }\href {\doibase 10.1209/epl/i2005-10471-9} {\bibfield
  {journal} {\bibinfo  {journal} {Europhys. Lett.}\ }\textbf {\bibinfo {volume}
  {73}},\ \bibinfo {pages} {892} (\bibinfo {year} {2006})}\BibitemShut
  {NoStop}%
\bibitem [{\citenamefont {Grigera}\ \emph {et~al.}(2003)\citenamefont
  {Grigera}, \citenamefont {Mart{\'{i}}n-Mayor}, \citenamefont {Parisi},\ and\
  \citenamefont {Verrocchio}}]{Grigera2003PhononLiquids}%
  \BibitemOpen
  \bibfield  {author} {\bibinfo {author} {\bibfnamefont {T.~S.}\ \bibnamefont
  {Grigera}}, \bibinfo {author} {\bibfnamefont {V.}~\bibnamefont
  {Mart{\'{i}}n-Mayor}}, \bibinfo {author} {\bibfnamefont {G.}~\bibnamefont
  {Parisi}}, \ and\ \bibinfo {author} {\bibfnamefont {P.}~\bibnamefont
  {Verrocchio}},\ }\href {\doibase 10.1038/nature01475} {\bibfield  {journal}
  {\bibinfo  {journal} {Nature}\ }\textbf {\bibinfo {volume} {422}},\ \bibinfo
  {pages} {289} (\bibinfo {year} {2003})}\BibitemShut {NoStop}%
\bibitem [{\citenamefont {Il'in}\ \emph {et~al.}(1987)\citenamefont {Il'in},
  \citenamefont {Karpov},\ and\ \citenamefont {Parshin}}]{ilyin1987parameters}%
  \BibitemOpen
  \bibfield  {author} {\bibinfo {author} {\bibfnamefont {M.~A.}\ \bibnamefont
  {Il'in}}, \bibinfo {author} {\bibfnamefont {V.~G.}\ \bibnamefont {Karpov}}, \
  and\ \bibinfo {author} {\bibfnamefont {D.~A.}\ \bibnamefont {Parshin}},\
  }\href {http://www.jetp.ac.ru/cgi-bin/dn/e_065_01_0165.pdf} {\bibfield
  {journal} {\bibinfo  {journal} {Zh. Eksp. Teor. Fiz.}\ }\textbf {\bibinfo
  {volume} {92}},\ \bibinfo {pages} {291} (\bibinfo {year} {1987})}\BibitemShut
  {NoStop}%
\bibitem [{\citenamefont {Parshin}(1993)}]{Parshin1993SoftGlasses}%
  \BibitemOpen
  \bibfield  {author} {\bibinfo {author} {\bibfnamefont {D.~A.}\ \bibnamefont
  {Parshin}},\ }\href {\doibase 10.1088/0031-8949/1993/T49A/030} {\bibfield
  {journal} {\bibinfo  {journal} {Phys. Scr.}\ }\textbf {\bibinfo {volume}
  {T49A}},\ \bibinfo {pages} {180} (\bibinfo {year} {1993})}\BibitemShut
  {NoStop}%
\bibitem [{\citenamefont {Buchenau}\ \emph {et~al.}(1991)\citenamefont
  {Buchenau}, \citenamefont {Galperin}, \citenamefont {Gurevich},\ and\
  \citenamefont {Schober}}]{Buchenau1991AnharmonicGlasses}%
  \BibitemOpen
  \bibfield  {author} {\bibinfo {author} {\bibfnamefont {U.}~\bibnamefont
  {Buchenau}}, \bibinfo {author} {\bibfnamefont {Y.~M.}\ \bibnamefont
  {Galperin}}, \bibinfo {author} {\bibfnamefont {V.~L.}\ \bibnamefont
  {Gurevich}}, \ and\ \bibinfo {author} {\bibfnamefont {H.~R.}\ \bibnamefont
  {Schober}},\ }\href {\doibase 10.1360/zd-2013-43-6-1064} {\bibfield
  {journal} {\bibinfo  {journal} {Phys. Rev. B}\ }\textbf {\bibinfo {volume}
  {43}},\ \bibinfo {pages} {5039} (\bibinfo {year} {1991})}\BibitemShut
  {NoStop}%
\bibitem [{\citenamefont {Ramos}\ and\ \citenamefont
  {Bachenau}(1998)}]{Ramos1998BeyondModel}%
  \BibitemOpen
  \bibfield  {author} {\bibinfo {author} {\bibfnamefont {M.~A.}\ \bibnamefont
  {Ramos}}\ and\ \bibinfo {author} {\bibfnamefont {U.}~\bibnamefont
  {Bachenau}},\ }in\ \href@noop {} {\emph {\bibinfo {booktitle} {Tunneling
  Systems in Amorphous and Crystalline Solids}}},\ \bibinfo {editor} {edited
  by\ \bibinfo {editor} {\bibfnamefont {P.}~\bibnamefont {Esquinazi}}}\
  (\bibinfo  {publisher} {Springer},\ \bibinfo {year} {1998})\ pp.\ \bibinfo
  {pages} {527--569}\BibitemShut {NoStop}%
\bibitem [{\citenamefont {Ramos}(2004)}]{Ramos2004AreDifferent}%
  \BibitemOpen
  \bibfield  {author} {\bibinfo {author} {\bibfnamefont {M.~A.}\ \bibnamefont
  {Ramos}},\ }\href {\doibase 10.1080/14786430310001644053} {\bibfield
  {journal} {\bibinfo  {journal} {Philos. Mag.}\ }\textbf {\bibinfo {volume}
  {84}},\ \bibinfo {pages} {1313} (\bibinfo {year} {2004})}\BibitemShut
  {NoStop}%
\bibitem [{\citenamefont {Mahan}\ \emph {et~al.}(1991)\citenamefont {Mahan},
  \citenamefont {Nelson}, \citenamefont {Salamon},\ and\ \citenamefont
  {Crandall}}]{Mahan1991}%
  \BibitemOpen
  \bibfield  {author} {\bibinfo {author} {\bibfnamefont {A.~H.}\ \bibnamefont
  {Mahan}}, \bibinfo {author} {\bibfnamefont {B.~P.}\ \bibnamefont {Nelson}},
  \bibinfo {author} {\bibfnamefont {S.}~\bibnamefont {Salamon}}, \ and\
  \bibinfo {author} {\bibfnamefont {R.~S.}\ \bibnamefont {Crandall}},\ }\href
  {https://doi.org/10.1557/PROC-219-673} {\bibfield  {journal} {\bibinfo
  {journal} {J. Non-Cryst. Solids}\ }\textbf {\bibinfo {volume} {137{\&}138}},\
  \bibinfo {pages} {657} (\bibinfo {year} {1991})}\BibitemShut {NoStop}%
\bibitem [{\citenamefont {Queen}\ and\ \citenamefont
  {Hellman}(2009)}]{Queen2009}%
  \BibitemOpen
  \bibfield  {author} {\bibinfo {author} {\bibfnamefont {D.~R.}\ \bibnamefont
  {Queen}}\ and\ \bibinfo {author} {\bibfnamefont {F.}~\bibnamefont
  {Hellman}},\ }\href {\doibase 10.1063/1.3142463} {\bibfield  {journal}
  {\bibinfo  {journal} {Rev. Sci. Instrum.}\ }\textbf {\bibinfo {volume}
  {80}},\ \bibinfo {pages} {063901} (\bibinfo {year} {2009})}\BibitemShut
  {NoStop}%
\bibitem [{\citenamefont {Lee}\ \emph {et~al.}(2005)\citenamefont {Lee},
  \citenamefont {Ohmori}, \citenamefont {Shin}, \citenamefont {Cahill},
  \citenamefont {Petrov},\ and\ \citenamefont {Greene}}]{Lee2005}%
  \BibitemOpen
  \bibfield  {author} {\bibinfo {author} {\bibfnamefont {T.}~\bibnamefont
  {Lee}}, \bibinfo {author} {\bibfnamefont {K.}~\bibnamefont {Ohmori}},
  \bibinfo {author} {\bibfnamefont {C.-S.}\ \bibnamefont {Shin}}, \bibinfo
  {author} {\bibfnamefont {D.~G.}\ \bibnamefont {Cahill}}, \bibinfo {author}
  {\bibfnamefont {I.}~\bibnamefont {Petrov}}, \ and\ \bibinfo {author}
  {\bibfnamefont {J.~E.}\ \bibnamefont {Greene}},\ }\href {\doibase
  10.1103/PhysRevB.71.144106} {\bibfield  {journal} {\bibinfo  {journal} {Phys.
  Rev. B}\ }\textbf {\bibinfo {volume} {71}},\ \bibinfo {pages} {144106}
  (\bibinfo {year} {2005})}\BibitemShut {NoStop}%
\bibitem [{\citenamefont {Liu}\ \emph {et~al.}(2009)\citenamefont {Liu},
  \citenamefont {Feldman}, \citenamefont {Cahill}, \citenamefont {Crandall},
  \citenamefont {Bernstein}, \citenamefont {Photiadis}, \citenamefont {Mehl},\
  and\ \citenamefont {Papaconstantopoulos}}]{Liu2009}%
  \BibitemOpen
  \bibfield  {author} {\bibinfo {author} {\bibfnamefont {X.}~\bibnamefont
  {Liu}}, \bibinfo {author} {\bibfnamefont {J.~L.}\ \bibnamefont {Feldman}},
  \bibinfo {author} {\bibfnamefont {D.~G.}\ \bibnamefont {Cahill}}, \bibinfo
  {author} {\bibfnamefont {R.~S.}\ \bibnamefont {Crandall}}, \bibinfo {author}
  {\bibfnamefont {N.}~\bibnamefont {Bernstein}}, \bibinfo {author}
  {\bibfnamefont {D.~M.}\ \bibnamefont {Photiadis}}, \bibinfo {author}
  {\bibfnamefont {M.~J.}\ \bibnamefont {Mehl}}, \ and\ \bibinfo {author}
  {\bibfnamefont {D.~A.}\ \bibnamefont {Papaconstantopoulos}},\ }\href
  {\doibase 10.1103/PhysRevLett.102.035901} {\bibfield  {journal} {\bibinfo
  {journal} {Phys. Rev. Lett.}\ }\textbf {\bibinfo {volume} {102}},\ \bibinfo
  {pages} {035901} (\bibinfo {year} {2009})}\BibitemShut {NoStop}%
\bibitem [{\citenamefont {Street}(1991{\natexlab{a}})}]{Street1991}%
  \BibitemOpen
  \bibfield  {author} {\bibinfo {author} {\bibfnamefont {R.~A.}\ \bibnamefont
  {Street}},\ }in\ \href {\doibase 10.1002/adma.19920040419} {\emph {\bibinfo
  {booktitle} {Hydrogenated amorphous silicon}}},\ \bibinfo {editor} {edited
  by\ \bibinfo {editor} {\bibfnamefont {R.~W.}\ \bibnamefont {Cahn}}, \bibinfo
  {editor} {\bibfnamefont {E.~A.}\ \bibnamefont {Davis}}, \ and\ \bibinfo
  {editor} {\bibfnamefont {I.~M.}\ \bibnamefont {Ward}}}\ (\bibinfo
  {publisher} {Cambridge University Press},\ \bibinfo {year} {1991})\
  Chap.~\bibinfo {chapter} {6}, pp.\ \bibinfo {pages} {169--223}\BibitemShut
  {NoStop}%
\bibitem [{\citenamefont {Berret}\ and\ \citenamefont
  {Meissner}(1988)}]{Berret1988}%
  \BibitemOpen
  \bibfield  {author} {\bibinfo {author} {\bibfnamefont {J.~F.}\ \bibnamefont
  {Berret}}\ and\ \bibinfo {author} {\bibfnamefont {M.}~\bibnamefont
  {Meissner}},\ }\href {https://doi.org/10.1007/BF01320540} {\bibfield
  {journal} {\bibinfo  {journal} {Z. Phys. B}\ }\textbf {\bibinfo {volume}
  {70}},\ \bibinfo {pages} {65} (\bibinfo {year} {1988})}\BibitemShut {NoStop}%
\bibitem [{\citenamefont {Queen}\ \emph
  {et~al.}(2015{\natexlab{b}})\citenamefont {Queen}, \citenamefont {Liu},
  \citenamefont {Karel}, \citenamefont {Wang}, \citenamefont {Crandall},
  \citenamefont {Metcalf},\ and\ \citenamefont {Hellman}}]{Queen2015EPL}%
  \BibitemOpen
  \bibfield  {author} {\bibinfo {author} {\bibfnamefont {D.~R.}\ \bibnamefont
  {Queen}}, \bibinfo {author} {\bibfnamefont {X.}~\bibnamefont {Liu}}, \bibinfo
  {author} {\bibfnamefont {J.}~\bibnamefont {Karel}}, \bibinfo {author}
  {\bibfnamefont {Q.}~\bibnamefont {Wang}}, \bibinfo {author} {\bibfnamefont
  {R.~S.}\ \bibnamefont {Crandall}}, \bibinfo {author} {\bibfnamefont {T.~H.}\
  \bibnamefont {Metcalf}}, \ and\ \bibinfo {author} {\bibfnamefont
  {F.}~\bibnamefont {Hellman}},\ }\href {\doibase 10.1209/0295-5075/112/26001}
  {\bibfield  {journal} {\bibinfo  {journal} {Europhys. Lett.}\ }\textbf
  {\bibinfo {volume} {112}},\ \bibinfo {pages} {26001} (\bibinfo {year}
  {2015}{\natexlab{b}})}\BibitemShut {NoStop}%
\bibitem [{\citenamefont {Stephens}(1973)}]{Stephens1973}%
  \BibitemOpen
  \bibfield  {author} {\bibinfo {author} {\bibfnamefont {R.~B.}\ \bibnamefont
  {Stephens}},\ }\href {https://doi.org/10.1103/PhysRevB.8.2896} {\bibfield
  {journal} {\bibinfo  {journal} {Phys. Rev. B}\ }\textbf {\bibinfo {volume}
  {8}},\ \bibinfo {pages} {2896} (\bibinfo {year} {1973})}\BibitemShut
  {NoStop}%
\bibitem [{\citenamefont {Nakhmanson}\ and\ \citenamefont
  {Drabold}(2000{\natexlab{a}})}]{Nakhmanson2000}%
  \BibitemOpen
  \bibfield  {author} {\bibinfo {author} {\bibfnamefont {S.~M.}\ \bibnamefont
  {Nakhmanson}}\ and\ \bibinfo {author} {\bibfnamefont {D.~A.}\ \bibnamefont
  {Drabold}},\ }\href {\doibase 10.1103/PhysRevB.61.5376} {\bibfield  {journal}
  {\bibinfo  {journal} {Phys. Rev. B}\ }\textbf {\bibinfo {volume} {61}},\
  \bibinfo {pages} {5376} (\bibinfo {year} {2000}{\natexlab{a}})}\BibitemShut
  {NoStop}%
\bibitem [{\citenamefont {Rau}\ \emph {et~al.}(1995)\citenamefont {Rau},
  \citenamefont {Enss}, \citenamefont {Hunklinger}, \citenamefont {Neu},\ and\
  \citenamefont {W{\"{u}}rger}}]{Rau1995}%
  \BibitemOpen
  \bibfield  {author} {\bibinfo {author} {\bibfnamefont {S.}~\bibnamefont
  {Rau}}, \bibinfo {author} {\bibfnamefont {C.}~\bibnamefont {Enss}}, \bibinfo
  {author} {\bibfnamefont {S.}~\bibnamefont {Hunklinger}}, \bibinfo {author}
  {\bibfnamefont {P.}~\bibnamefont {Neu}}, \ and\ \bibinfo {author}
  {\bibfnamefont {A.}~\bibnamefont {W{\"{u}}rger}},\ }\href {\doibase
  10.1103/PhysRevB.52.7179} {\bibfield  {journal} {\bibinfo  {journal} {Phys.
  Rev. B}\ }\textbf {\bibinfo {volume} {52}},\ \bibinfo {pages} {7179}
  (\bibinfo {year} {1995})}\BibitemShut {NoStop}%
\bibitem [{\citenamefont {Ediger}(2017)}]{Ediger2017Perspective:Glasses}%
  \BibitemOpen
  \bibfield  {author} {\bibinfo {author} {\bibfnamefont {M.~D.}\ \bibnamefont
  {Ediger}},\ }\href {\doibase 10.1063/1.5006265} {\bibfield  {journal}
  {\bibinfo  {journal} {J. Chem. Phys.}\ }\textbf {\bibinfo {volume} {147}},\
  \bibinfo {pages} {210901} (\bibinfo {year} {2017})}\BibitemShut {NoStop}%
\bibitem [{\citenamefont {Wu}\ \emph {et~al.}(1996)\citenamefont {Wu},
  \citenamefont {Stephen}, \citenamefont {Han}, \citenamefont {Rutland},
  \citenamefont {Crandall},\ and\ \citenamefont {Mahan}}]{Wu1996}%
  \BibitemOpen
  \bibfield  {author} {\bibinfo {author} {\bibfnamefont {Y.}~\bibnamefont
  {Wu}}, \bibinfo {author} {\bibfnamefont {J.~T.}\ \bibnamefont {Stephen}},
  \bibinfo {author} {\bibfnamefont {D.~X.}\ \bibnamefont {Han}}, \bibinfo
  {author} {\bibfnamefont {J.~M.}\ \bibnamefont {Rutland}}, \bibinfo {author}
  {\bibfnamefont {R.~S.}\ \bibnamefont {Crandall}}, \ and\ \bibinfo {author}
  {\bibfnamefont {A.~H.}\ \bibnamefont {Mahan}},\ }\href {\doibase
  10.1103/PhysRevLett.77.2049} {\bibfield  {journal} {\bibinfo  {journal}
  {Phys. Rev. Lett.}\ }\textbf {\bibinfo {volume} {77}},\ \bibinfo {pages}
  {2049} (\bibinfo {year} {1996})}\BibitemShut {NoStop}%
\bibitem [{\citenamefont {Su}\ \emph {et~al.}(2000)\citenamefont {Su},
  \citenamefont {Chen}, \citenamefont {Taylor}, \citenamefont {Crandall},\ and\
  \citenamefont {Mahan}}]{Su2000}%
  \BibitemOpen
  \bibfield  {author} {\bibinfo {author} {\bibfnamefont {T.}~\bibnamefont
  {Su}}, \bibinfo {author} {\bibfnamefont {S.}~\bibnamefont {Chen}}, \bibinfo
  {author} {\bibfnamefont {P.~C.}\ \bibnamefont {Taylor}}, \bibinfo {author}
  {\bibfnamefont {R.~S.}\ \bibnamefont {Crandall}}, \ and\ \bibinfo {author}
  {\bibfnamefont {A.~H.}\ \bibnamefont {Mahan}},\ }\href
  {https://doi.org/10.1103/PhysRevB.62.12849} {\bibfield  {journal} {\bibinfo
  {journal} {Phys. Rev. B}\ }\textbf {\bibinfo {volume} {62}},\ \bibinfo
  {pages} {12849} (\bibinfo {year} {2000})}\BibitemShut {NoStop}%
\bibitem [{\citenamefont {Reimer}\ \emph {et~al.}(1981)\citenamefont {Reimer},
  \citenamefont {Vaughan},\ and\ \citenamefont {Knights}}]{Reimer1981}%
  \BibitemOpen
  \bibfield  {author} {\bibinfo {author} {\bibfnamefont {J.~A.}\ \bibnamefont
  {Reimer}}, \bibinfo {author} {\bibfnamefont {R.~W.}\ \bibnamefont {Vaughan}},
  \ and\ \bibinfo {author} {\bibfnamefont {J.~C.}\ \bibnamefont {Knights}},\
  }\href {https://doi.org/10.1016/0038-1098(81)90734-1} {\bibfield  {journal}
  {\bibinfo  {journal} {Solid State Commun.}\ }\textbf {\bibinfo {volume}
  {37}},\ \bibinfo {pages} {161} (\bibinfo {year} {1981})}\BibitemShut
  {NoStop}%
\bibitem [{\citenamefont {Baugh}\ \emph {et~al.}(1999)\citenamefont {Baugh},
  \citenamefont {Han}, \citenamefont {Wang},\ and\ \citenamefont
  {Wu}}]{Baugh1999}%
  \BibitemOpen
  \bibfield  {author} {\bibinfo {author} {\bibfnamefont {J.}~\bibnamefont
  {Baugh}}, \bibinfo {author} {\bibfnamefont {D.}~\bibnamefont {Han}}, \bibinfo
  {author} {\bibfnamefont {Q.}~\bibnamefont {Wang}}, \ and\ \bibinfo {author}
  {\bibfnamefont {Y.}~\bibnamefont {Wu}},\ }\href {\doibase
  10.1557/PROC-557-383} {\bibfield  {journal} {\bibinfo  {journal} {Mat. Res.
  Soc. Symp. Proc.}\ }\textbf {\bibinfo {volume} {557}},\ \bibinfo {pages}
  {383} (\bibinfo {year} {1999})}\BibitemShut {NoStop}%
\bibitem [{\citenamefont {Beyer}\ and\ \citenamefont
  {Wagner}(1983)}]{Beyer1983}%
  \BibitemOpen
  \bibfield  {author} {\bibinfo {author} {\bibfnamefont {W.}~\bibnamefont
  {Beyer}}\ and\ \bibinfo {author} {\bibfnamefont {H.}~\bibnamefont {Wagner}},\
  }\href {https://doi.org/10.1016/0022-3093(83)90547-1} {\bibfield  {journal}
  {\bibinfo  {journal} {J. Non-Cryst. Solids}\ }\textbf {\bibinfo {volume}
  {59{\&}60}},\ \bibinfo {pages} {161} (\bibinfo {year} {1983})}\BibitemShut
  {NoStop}%
\bibitem [{\citenamefont {Branz}(1998)}]{Branz1998}%
  \BibitemOpen
  \bibfield  {author} {\bibinfo {author} {\bibfnamefont {H.~M.}\ \bibnamefont
  {Branz}},\ }\href {https://doi.org/10.1016/S0038-1098(97)10142-9} {\bibfield
  {journal} {\bibinfo  {journal} {Solid State Commun.}\ }\textbf {\bibinfo
  {volume} {105}},\ \bibinfo {pages} {387} (\bibinfo {year}
  {1998})}\BibitemShut {NoStop}%
\bibitem [{\citenamefont {Fujita}\ \emph {et~al.}(1994)\citenamefont {Fujita},
  \citenamefont {Amaguchi},\ and\ \citenamefont
  {Morigaki}}]{Fujita1994KineticsSilicon}%
  \BibitemOpen
  \bibfield  {author} {\bibinfo {author} {\bibfnamefont {Y.}~\bibnamefont
  {Fujita}}, \bibinfo {author} {\bibfnamefont {M.}~\bibnamefont {Amaguchi}}, \
  and\ \bibinfo {author} {\bibfnamefont {K.}~\bibnamefont {Morigaki}},\ }\href
  {\doibase 10.1080/13642819408236879} {\bibfield  {journal} {\bibinfo
  {journal} {Philos. Mag. B}\ }\textbf {\bibinfo {volume} {69}},\ \bibinfo
  {pages} {57} (\bibinfo {year} {1994})}\BibitemShut {NoStop}%
\bibitem [{\citenamefont {Graebner}\ \emph {et~al.}(1985)\citenamefont
  {Graebner}, \citenamefont {Allen},\ and\ \citenamefont
  {Golding}}]{Graebner1985}%
  \BibitemOpen
  \bibfield  {author} {\bibinfo {author} {\bibfnamefont {J.~E.}\ \bibnamefont
  {Graebner}}, \bibinfo {author} {\bibfnamefont {L.~C.}\ \bibnamefont {Allen}},
  \ and\ \bibinfo {author} {\bibfnamefont {B.}~\bibnamefont {Golding}},\ }\href
  {https://doi.org/10.1103/PhysRevB.31.904} {\bibfield  {journal} {\bibinfo
  {journal} {Phys. Rev. B}\ }\textbf {\bibinfo {volume} {31}},\ \bibinfo
  {pages} {904} (\bibinfo {year} {1985})}\BibitemShut {NoStop}%
\bibitem [{\citenamefont {Street}(1991{\natexlab{b}})}]{Street1991a}%
  \BibitemOpen
  \bibfield  {author} {\bibinfo {author} {\bibfnamefont {R.~A.}\ \bibnamefont
  {Street}},\ }\href {http://adsabs.harvard.edu/abs/1991PhRvB..43.2454S}
  {\bibfield  {journal} {\bibinfo  {journal} {Phys. Rev. B}\ }\textbf {\bibinfo
  {volume} {43}},\ \bibinfo {pages} {2454} (\bibinfo {year}
  {1991}{\natexlab{b}})}\BibitemShut {NoStop}%
\bibitem [{\citenamefont {Biswas}\ and\ \citenamefont {Li}(1999)}]{Biswas1999}%
  \BibitemOpen
  \bibfield  {author} {\bibinfo {author} {\bibfnamefont {R.}~\bibnamefont
  {Biswas}}\ and\ \bibinfo {author} {\bibfnamefont {Y.-P.}\ \bibnamefont
  {Li}},\ }\href {https://doi.org/10.1103/PhysRevLett.82.2512} {\bibfield
  {journal} {\bibinfo  {journal} {Phys. Rev. Lett.}\ }\textbf {\bibinfo
  {volume} {82}},\ \bibinfo {pages} {2512} (\bibinfo {year}
  {1999})}\BibitemShut {NoStop}%
\bibitem [{\citenamefont {Nakhmanson}\ and\ \citenamefont
  {Drabold}(2000{\natexlab{b}})}]{Nakhmanson2000a}%
  \BibitemOpen
  \bibfield  {author} {\bibinfo {author} {\bibfnamefont {S.~M.}\ \bibnamefont
  {Nakhmanson}}\ and\ \bibinfo {author} {\bibfnamefont {D.~A.}\ \bibnamefont
  {Drabold}},\ }\href {\doibase 10.1016/S0022-3093(99)00795-4} {\bibfield
  {journal} {\bibinfo  {journal} {J. Non-Cryst. Solids}\ }\textbf {\bibinfo
  {volume} {266-269}},\ \bibinfo {pages} {156} (\bibinfo {year}
  {2000}{\natexlab{b}})}\BibitemShut {NoStop}%
\end{thebibliography}%
\end{document}